# ACCELERATOR UPGRADES REQUIRED FOR BEAM OPERATION AT FERMILAB IN THE PIP-II/DUNE ERA*

S. Krishnagopal†, M. Convery, J. Dey, Fermi National Accelerator Laboratory, Batavia, United States

*Abstract*

The Proton Improvement Plan - II (PIP-II) injector linac is an 800 MeV superconducting $H^-$ linac, christened Linac2, that will replace the existing 400 MeV injector to the accelerator complex at Fermilab. The higher energy, intensity and repetition rate require various upgrades to the existing accelerator complex consisting of the Booster, the Recycler Ring and the Main Injector, in order to be able to accept and accelerate beam from PIP-II. In this paper we discuss various upgrades that are required and steps being taken to implement them.

## INTRODUCTION

The Proton Improvement Plan-II (PIP-II) [1] will deliver an 800 MeV superconducting $H^-$ linac, Linac2, that will replace the existing 400 MeV normal-conducting injector. PIP-II/Linac2 is an essential upgrade of the existing Fermilab Accelerator Complex that will provide a 1.2 MW proton beam for the Long-Baseline Neutrino Facility and the Deep Underground Neutrino Experiment (LBNF/DUNE) [2].

The higher injection energy, repetition rate and beam intensity provided by Linac2 necessitate a number of changes and upgrades to the rest of the Accelerator Complex comprising three synchrotrons – the Booster, the Recycler Ring (RR) and the Main Injector (MI). Other upgrades are also necessitated by reliability and end-of-life issues of various components, given the vintage of the synchrotrons (between 30 and 50 years). These have been the subject of considerable study over the years by the Accelerator Complex Studies Task Force, and the main conclusions were summarized in the 2024 AD Preparation for the DUNE-PIP-II Era Workshop [3]. In this paper we summarize the drivers of change, and describe the scope of the changes that are addressed by just two complementary efforts: the Accelerator Upgrades sub-project of PIP-II, and Accelerator Complex Evolution (ACE) – PIP-II Integration (P2I) program, that together ensure a smooth transition of the Accelerator Complex into the DUNE/PIP-II era.

## DRIVERS OF CHANGE

The key Linac2 beam parameters that drive upgrades across the existing Accelerator Complex are summarized in Table 1. The injection energy doubles from 400 to 800 MeV, the beam intensity rises from $4.5\times10^{12}$ to $6.7\times10^{12}$ protons/pulse, the repetition rate increases from 15 to 20 Hz, and the injected pulse length grows from 30 μs to 550 μs. In addition, beam will be injected into a new section of the Booster – period 11 instead of period 1.

Table 1: Key Linac2 Beam Parameters vs. the existing 400 MeV linac.

| Parameter | Existing-linac | Linac2 |
|---|---|---|
| Beam injection energy (MeV) | 400 | 800 |
| Beam intensity (protons/pulse) | $4.5\times10^{12}$ | $6.7\times10^{12}$ |
| Repetition rate (Hz) | 15 | 20 |
| Injection pulse length (μs) | 30 | 550 |
| Booster injection area | L1 | L11 |

Clearly these parameters will drive a large number of changes and upgrades, not all of which can be discussed in this paper. Here, as an example, are a few of the impacts; greater details can be found in Ref. 3.

***(1) Injection:***
- (a) Shorter combined function magnets (CFM) to make space for injection.
- (b) New ORBUMP and painting magnets for the actual injection.

***(2) Repetition rate:***
- (a) All CFM girder capacitors changed for 20 Hz frequency.
- (b) Much of the electronics (BPMs, etc.) upgraded for 20 Hz operation.
- (c) More heating of components, requiring more cooling (water and air).

***(3) Intensity:***
- (a) Need more RF voltage in the Booster and more RF power in MI.
- (b) New transverse and longitudinal dampers to address intensity-driven instabilities.
- (c) New Gamma-T jump system for the MI.
- (d) New Gamma-T jump system for the Booster.

From this example, 1, 2(a) and 3(a-c) are addressed by the PIP-II project, while 2(b) and 3(d) are addressed by ACE-P2I, which also addresses some spares and end-of-life issues. Item 2(c), and other infrastructure related impacts, are addressed by the Accelerator Infrastructure

†skrishna@fnal.gov

Readiness (AIR) component of ACE, which is beyond the scope of this paper.

## PIP-II ACCELERATOR UPGRADES

The Accelerator Upgrades (AccU) Level 2 of the PIP-II project addresses some of upgrades required for the Booster and the Main Injector.

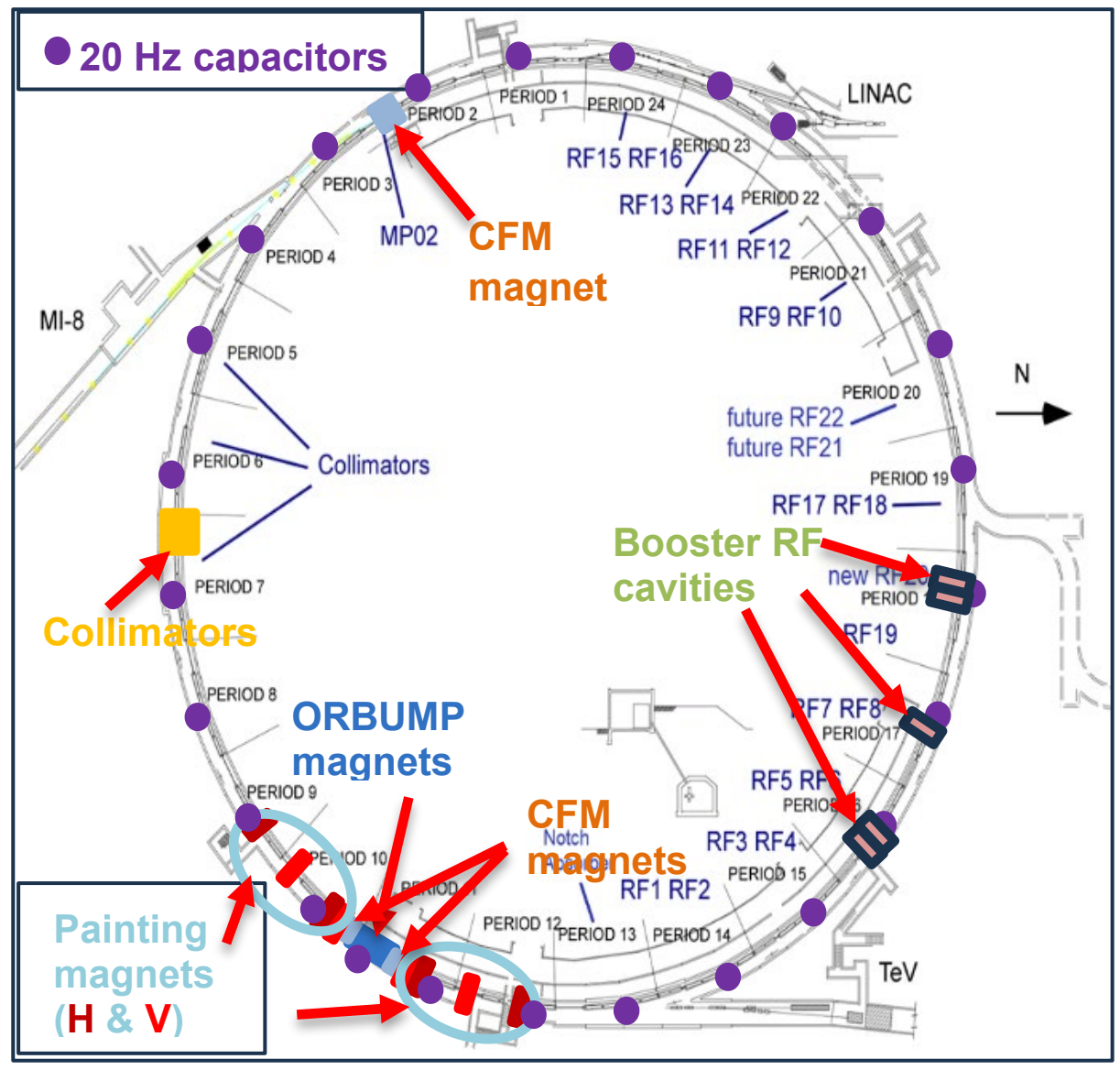

Fig. 1: PIP-II AccU scope in the Booster.

The AccU scope in the Booster is shown in Fig. 1. The major impact is on the new injection region. To gain around 1 m in the straight section in period 11 for injection from PIP-II, it has been necessary to redesign two of the CFM magnets to have a shorter length (reduced from 3.2 m to 2.4 m). This also necessitates two new CFM girders. For the injection itself, there are four new 'ORBUMP' magnets to bump the beam into the Booster, sitting on a new injection girder – which also has a foil system to strip the $H^-$ ions to inject protons. Because of the higher intensity of the Linac2 beam, there are eight new painting magnets for transverse phase-space painting. There is also a 200 W injection absorber for unstripped ions and neutrals.

Beyond the injection region, there are two new beam collimators, five new Booster cavities (out of 22) with a larger aperture and gradient to deliver the required voltage, a new and wider CFM magnet at extraction, to reduce beam loss, and new capacitors on all girders around the ring to resonate at the new repetition rate of 20 Hz.

The AccU scope in the Main Injector is shown in Fig. 2. Each of the 20 MI cavities presently has one power amplifier installed on the cavity, and the upgrade will install a second amplifier on each cavity to provide the required power for beam acceleration and beam loading. This also necessitates adding a second power amplifier to the modulators and also redesigning the modulator cabinet and components to be more compact so that all the hardware fits in the same footprint.

The higher intensity in the MI means that there will be greater beam loss during transition crossing. Therefore, a new Gamma-T transition system will be installed in the MI, comprising four sets of four quadrupole magnets located around the ring; each set of four magnets will have a new power supply.

All these components are either in advanced stages of design and proceeding to procurement or are already under development. Installation of these components will be done by the Accelerator Directorate (AD). All installation is expected to be completed by the end of 2029, before Linac2 beam is injected into the Booster.

The scope boundary between the PIP-II Project and the Accelerator Directorate (AD) has been captured in the PIP-II to Accelerator Complex Interface Document [4] and a set of bidirectional hand-off milestones has been incorporated into the PIP-II P6 schedule. AD will develop a schedule for their installation activities These activities and milestones will be reviewed regularly and tracked in P6.

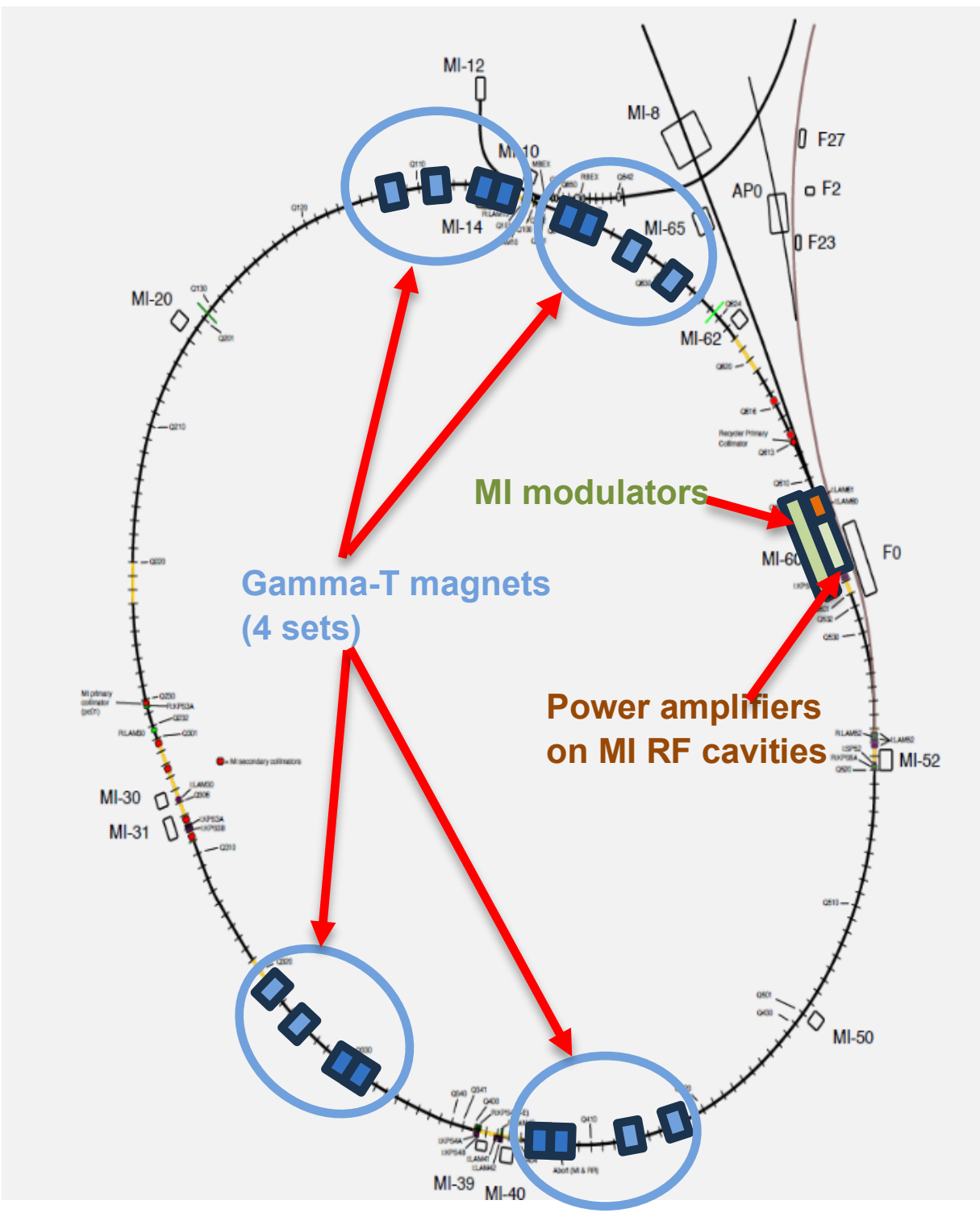

Fig. 2: PIP-II AccU scope in the Main Injector (MI).

## ACE–PIP-II INTEGRATION (P2I)

Several items that are necessary for operation in the PIP-II era — but that are not required to meet the PIP-II Project Key Performance Parameters (KPPs) — fall outside the scope of the PIP-II project. These are collected in the P2I sub-program within the Accelerator Complex Evolution (ACE) program and are funded through Accelerator Operations. Table 2 summarizes these activities and their durations.

Table 2: ACE-P2I activities and their durations.

| | FY26 | FY27 | FY28 | FY29 | FY30 | FY31 | FY32 | FY33 |
|---|---|---|---|---|---|---|---|---|
| PIP-II related infrastructure | ■ | ■ | ■ | | | | | |
| 20 Hz instrumentation | | ■ | ■ | ■ | ■ | | | |
| Booster RF cavities & tuners | | ■ | ■ | ■ | ■ | ■ | ■ | ■ |
| BTL buncher | | | ■ | ■ | | | | |
| Power supplies | | | ■ | ■ | ■ | ■ | ■ | ■ |
| HWR spare | | | ■ | ■ | ■ | ■ | ■ | ■ |
| Booster Gamma-T | | | | | | ■ | ■ | |

- **PIP-II-related infrastructure.** This covers a number of activities related to PIP-II-related accelerator infrastructure upgrades that need to be completed before Linac2 commissioning. These include upgrades to the power supply for the CFM magnets for 20 Hz operation, RF test stands for testing couplers, upgrade to some beam position monitors, etc.
- **20 Hz instrumentation.** Almost all the instrumentation used will need their electronics to be upgraded to work at 20 Hz.
- **Booster RF cavities & tuners.** To account for the higher intensity of the Linac2 beam in the Booster, the total voltage required needs to be increased. In addition, the Booster cavities are almost 50 years old and have reliability issues. While the PIP-II project will be building 5 new cavities with a larger bore, under P2I a further 12 cavities will be developed – 6 for operation and 6 for reliability.
- **BTL buncher.** As the Linac2 beam is transported from the exit of the linac to the entrance of the Booster, a distance of around 300 m, its energy spread increases which would result in unacceptable beam loss at injection during 1.2 MW operation. It is proposed to install two 9-cell buncher cavities to tune the longitudinal phase-space.
- **Power supplies.** Various power supplies in the Booster and MI need to be replaced to mitigate end-of-life risks. These include all the ferrite-bias supplies for tuning the RF cavities, all the ion-pump power supplies and controllers, and various critical magnet power supplies.
- **HWR spare.** The Half Wave Resonator (HWR) is the first cryomodule in the linac, accelerating beam from 2.1 to 10 MeV. HWR failure would be catastrophic since beam could not be successfully accelerated to 800 MeV. A spare HWR is not in the scope of the PIP-II project and is therefore included in ACE-P2I.
- **Booster Gamma**-T. At the full beam intensity of $6.7 \times 10^{12}$ protons/pulse (corresponding to 1.2 MW beam power), there would be unacceptable beam loss in the Booster during transition crossing [5]. A gamma-T transition system will therefore be installed to mitigate this issue.
- **RFQ spare**. The Linac2 Radio-Frequency Quadrupole (RFQ) will accelerate beam from 30 keV to 2.1 MeV and is critical component because in case of failure beam cannot be accelerated to 800 MeV. A spare RFQ is not in the scope of the PIP-II project and is therefore included in ACE-P2I. It's proposed to allow for a year or two of RFQ operation to gather data about potential design changes to the RFQ. Therefore, work on the spare RFQ will be taken up after 2033 (and is hence not included in Table 2).

## SUMMARY

The 800 MeV Linac2, being developed by the PIP-II Project, replaces the present 400 MeV injector but also drives substantial changes across the rest of the Fermilab Accelerator Complex. Some of these changes, essential for injecting beam into the Booster and accelerating it to 120 GeV in the Main Injector, are addressed by the PIP-II project. Interfaces between the PIP-II project and the Accelerator Directorate have been documented and signed off (PIP-II docDB-4298) and hand-off milestones are being finalized in the P6 schedule.

Some other changes, related to operation at the full power of 1.2 MW, as well as those related to reliability, are addressed by the PIP-II Integration (P2I) component of the Accelerator Complex Evolution (ACE) initiative.

Clearly, there are many other challenges and more extensive upgrades needed to modernize the Fermilab Accelerator Complex and keep it operating in the DUNE/PIP-II era for decades. There are other components to ACE that are discussed at another talk at this conference [6]. In addition, there are other initiatives at Fermilab to address upgrades related to infrastructure, controls, etc. We continue to remain vigilant to ensure that no scope needed for the successful operation of the Accelerator Complex in the DUNE/PIP-II era is missed.

## ACKNOWLEDGEMENTS

We thank our colleagues in the PIP-II project and in AD for all the work they have done on which this paper reports.